\documentclass[prl,superscriptaddress,amsfonts,twocolumn,showpacs,floatfix]{revtex4}

\usepackage{graphicx,color}
\usepackage{amsmath}
\usepackage{amssymb}
\usepackage{bm}
\usepackage{ulem}

\usepackage{txfonts}

\def\Vec#1{\bm{#1}}
\def\Hc2{H_\mathrm{c2}}
\def\SRO{\mathrm{Sr}_2\mathrm{RuO}_4}
\def\Tc{T_\mathrm{c}}

\begin{document}

\title{Sharp magnetization jump at the first-order superconducting transition in Sr$_2$RuO$_4$}

\author{Shunichiro Kittaka}
\affiliation{Institute for Solid State Physics, University of Tokyo, Kashiwa, Chiba 277-8581, Japan}
\author{Akira Kasahara}
\affiliation{Institute for Solid State Physics, University of Tokyo, Kashiwa, Chiba 277-8581, Japan}
\author{Toshiro Sakakibara}
\affiliation{Institute for Solid State Physics, University of Tokyo, Kashiwa, Chiba 277-8581, Japan}
\author{Daisuke Shibata}
\affiliation{Department of Physics, Kyoto University, Kyoto 606-8502, Japan}
\author{Shingo Yonezawa}
\affiliation{Department of Physics, Kyoto University, Kyoto 606-8502, Japan}
\author{Yoshiteru Maeno}
\affiliation{Department of Physics, Kyoto University, Kyoto 606-8502, Japan}
\author{Kenichi Tenya}
\affiliation{Faculty of Education, Shinshu University, Nagano 310-8512, Japan}
\author{Kazushige Machida}
\affiliation{Department of Physics, Okayama University, Okayama 700-8530, Japan}

\date{\today}

\begin{abstract}
The magnetization and magnetic torque of a high-quality single crystal of $\SRO$ have been measured down to 0.1~K 
under precise control of the magnetic-field orientation. 
When the magnetic field is applied exactly parallel to the $ab$ plane,
a sharp magnetization jump $4\pi\delta M$ of $(0.74\pm0.15)$~G at the upper critical field $H_{{\rm c2},{ab}}\sim 15$~kOe with a field hysteresis of 100~Oe is observed at low temperatures,
evidencing a first-order superconducting-normal transition. 
A strong magnetic torque appearing when $H$ is slightly tilted away from the $ab$ plane confirms an intrinsic anisotropy 
$\varGamma\!=\!\xi_{a}/\xi_{c}$ of as large as 60 even at 100~mK, in contrast with the observed $H_{{\rm c2}}$ anisotropy of $\sim 20$. 
The present results raise fundamental issues in both the existing spin-triplet and spin-singlet scenarios, providing, in turn, crucial hints toward the resolution of the superconducting nature of $\SRO$.
\end{abstract}


\maketitle

Spin-triplet superconductors have recently become increasingly familiar, because several promising candidates have been discovered, 
including ferromagnetic and noncentrosymmetric superconductors (SCs).
In general, crucial evidence for spin-triplet pairing  is provided by an invariance of the spin susceptibility 
across the superconducting-normal (S-N) transition on cooling; spins of the triplet Cooper pairs can be easily polarized along the field direction perpendicular to the $\Vec{d}$ vector, because equal-spin pairs can be formed under Zeeman-split Fermi surfaces.
If such a configuration is available, the Pauli-paramagnetic effect (PPE) is absent.
This feature of triplet SCs admits a high upper critical field $\Hc2$ that is determined solely by the orbital effect.

In $\SRO$, nuclear-magnetic-resonance (NMR) Knight-shift~\cite{Ishida2008JPCS} and polarized-neutron scattering~\cite{Duffy2000PRL} experiments 
have provided accumulating experimental evidence for a spin-triplet pairing with a chiral-$p$-wave state $\Vec{d}\!\!=\!\!\varDelta\hat{z}(k_x+ik_y)$~\cite{Mackenzie2003RMP, Maeno2012JPSJ}.
In addition, an unusual \textit{increase} of the NMR Knight shift has been recently found in the superconducting state~\cite{Ishida2014}, 
which has been understood in the framework of equal-spin pairing states including the proposed chiral-$p$-wave state~\cite{Miyake2014JPSJ}.
Despite compelling evidence for equal-spin pairing,
the upper critical field $\Hc2$ of $\SRO$ is strongly suppressed at low temperatures for $H\!\parallel\!ab$~\cite{Maeno2012JPSJ,Kittaka2009PRB}, 
in a fashion very similar to the PPE in spin-singlet SCs.
Accordingly, the $\Hc2$ anisotropy $\varGamma_{H}\!=\!H_{{\rm c2},ab}/H_{{\rm c2},c}$, which has a large value of $\sim\!60$ near $T_{\rm c}$, considerably reduces to $\sim\!20$ at 0.1~K~\cite{Kittaka2009JPCS,Kittaka2009PRB}. 
The origin of the strongly $T$-dependent $\varGamma_{H}$ has remained unresolved.
A similar $\Hc2$ limiting has also been observed  
for UPt$_3$ in $H\!\parallel\!c$~\cite{Sauls1994AP,Kittaka2013JPSJ}, another long-standing candidate for a spin-triplet superconductor; 
this limiting appears to be incompatible with an invariant Knight shift~\cite{Tou1996PRL,Tou1998PRL}. 
Quite recently, an even more mysterious phenomenon has been found in $\SRO$ by the magnetocaloric effect~\cite{Yonezawa2013PRL} and specific-heat measurements~\cite{Yonezawa2014JPSJ};
the S-N transition at $\Hc2$ becomes of first order below about 0.8~K when the magnetic field is applied closely parallel to the $ab$ plane. 
The first-order transition (FOT) has been reported to be accompanied by an entropy release of $(10\pm 3)$\% of the normal-state value at 0.2~K.

To our knowledge, the FOT in the presence of a strong suppression of $\Hc2$ has only been predicted for spin-singlet SCs exhibiting a strong PPE~\cite{Sarma1963JPCS}, 
as is the case of a $d$-wave superconductor CeCoIn$_5$~\cite{Matsuda2007JPSJ,Bianchi2002PRL,Adachi2003PRB}, 
in which a distinct jump in the magnetization has been observed~\cite{Tayama2002PRB}. 
Plausibly, Ba$_x$K$_{1-x}$Fe$_2$As$_2$~\cite{Zocco2013PRL,Terashima2013PRB} may also exhibit this type of FOT,
although the specific-heat and magnetization jumps have not yet been clearly observed~\cite{Burger2013PRB,Kittaka2014JPSJ}. 
In sharp contrast, the origin of FOT in $\SRO$ has remained unidentified
because no PPE is expected in the basal plane for the anticipated chiral-$p$-wave order parameter.
Further experimental investigations are clearly needed to uncover its mechanism.

To this end, quantitative evaluation of the magnetization jump at FOT is of primary interest. 
Magnetization of $\SRO$ in the superconducting state was previously measured with a crystal of dimensions of $3\!\times\!3\!\times\!0.5$~mm$^3$ ($\Tc\!\!=\!\!1.42$~K)~\cite{Tenya2006JPSJ}. 
The result shows a two-step change of slope below $H_{\rm c2}$ at 0.14~K, which was interpreted as the occurrence of a different superconducting phase; no clear evidence of FOT was obtained.
In the present Rapid Communication, we succeed in detecting a sharp magnetization jump of as large as 0.74~G at the FOT at 0.1~K using an ultraclean sample. 
Moreover, we estimate the intrinsic anisotropy parameter $\varGamma\!=\!\xi_{a}/\xi_{c}$ from the analysis of the magnetization torque that appears when $H$ is slightly tilted away from the basal plane, and 
obtain a significantly large value $\varGamma \sim 60$ even at 0.1~K, confirming the anisotropy reported in Ref.~\onlinecite{Rastovski2013PRL} but this time on a thermodynamical basis.
This result implies a large in-plane orbital limiting field of 45~kOe at $T\!=\!0$, three times as large as the observed $H_{{\rm c2}, {ab}}$.

Magnetization $M$ was measured down to 0.1~K in a dilution refrigerator 
by using a high-resolution capacitively detected Faraday magnetometer~\cite{Sakakibara1994JJAP}. 
A magnetic field as well as a field gradient of 500~Oe/cm were applied parallel to the vertical ($z$ axis) direction.
A high-quality single crystal of $\SRO$ ($\Tc\!\!=\!\!1.50$~K) used in the present study was grown by a floating-zone method~\cite{Mao2000MRB}.
To avoid possible crystal inhomogeneity as well as a field distribution in the sample caused by the field gradient,
a tiny crystal with dimensions of roughly $1 \times 0.4 \times 0.3$~mm$^3$ (0.72~mg mass) was selected.
It was fixed on a stage of the capacitor transducer
so that the crystal [110] axis, the longest dimension of the sample shape, is positioned at $z\!\!=\!\!0$ nearly parallel to the horizontal ($x$ axis) direction. 
The capacitor transducer was mounted on a stage that can be tilted around the $x$ axis, whose tilting angle was precisely controlled from the top of the refrigerator insert [see the Supplemental Material~\cite{supp} (I) for details].
The fine tuning of the angle $\theta$ between a magnetic field and the crystal $ab$ plane was accomplished with an accuracy of better than $\pm 0.05$~deg.

\begin{figure}
\begin{center}
\includegraphics[width=3.2in]{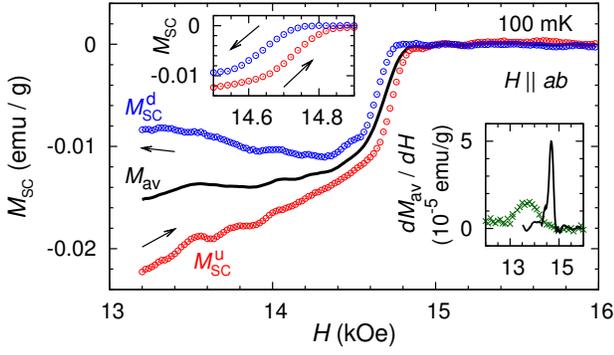}
\end{center}
\vspace{-0.2in}
\caption{
(Color online) 
Field dependence of the magnetization, $M_{\rm SC} = M - \chi_{\rm n}H$, at 0.1 K for $H\!\parallel\!ab$, 
where $\chi_{\rm n} H$ is the normal-state contribution. 
The solid line represents the $M_{\rm av}$ data obtained by averaging the increasing- and decreasing-field data ($M_{\rm SC}^{\rm u}$ and $M_{\rm SC}^{\rm d}$). 
The upper inset is an enlarged view near $\Hc2$. 
The lower inset shows $dM_{\rm av}/dH$, compared with the previous results~\cite{Tenya2006JPSJ} (crosses).
}
\label{MH}
\end{figure}

The field dependence of the superconducting magnetization $M_{\rm SC}\!=\!M\!-\!\chi_{\rm n}H$ measured at 0.1~K is shown in Fig.~\ref{MH}. 
Here, $\chi_{\rm n}$ is the paramagnetic susceptibility in the normal state. 
As clearly seen in the enlarged plot near $H_{\rm c2}$ (upper inset),
$M_{\rm SC}$ exhibits a sharp jump with a hysteresis of the onset field of about 100~Oe, clearly evidencing FOT.
Note that this hysteresis in the \textit{onset} field is totally different from the ordinary magnetization hysteresis caused by vortex pinning.
This magnetization jump grows below about 0.6~K [see the Supplemental material~\cite{supp} (II)].
The solid line in Fig.~\ref{MH} is the average of $M_{\rm SC}$ in the increasing and decreasing field sweeps, labeled as $M_{\rm av}$.
The lower inset of Fig.~\ref{MH} shows a field derivative $dM_{\rm av}/dH$ of the present data (solid line), indicating a sharp peak associated with the FOT at $H_{\rm c2}$.  
For comparison, $dM_{\rm av}/dH$ of the previous report~\cite{Tenya2006JPSJ} obtained with a field gradient of 800~Oe/cm is also shown (crosses).
The much narrower (larger) peak width (height) of the present result clearly demonstrates the higher quality of the present sample and smaller field inhomogeneity.

The data in Fig.~\ref{MH} show that the $M_{\rm av}$ jump at the first-order S-N transition, $\delta M$, is  $(0.01 \pm 0.002)$~emu/g, 
i.e., $4\pi\delta M\!=\!(0.74\pm0.15)$~G using a density of 5.9 g/cm${}^3$.
According to the Clausius-Clapeyron equation, $d\Hc2/dT\!=\!-\delta S/\delta M$,
$\delta M$ is estimated to be $(0.011 \!\pm\! 0.006)$~emu/g by using 
the previously-reported entropy jump $\delta S/T \!\!=\!\!(3.5 \pm 1)$~mJ/(K$^2$ mol) and $d\Hc2/dT\!\!\sim\!\!(-2\pm0.5)$~kOe/K~\cite{Yonezawa2013PRL} at 0.2~K.
Thus, the $\delta M$ value determined in the present experiment is consistent with the results of the thermal measurements~\cite{Yonezawa2013PRL,Yonezawa2014JPSJ}.

\begin{figure}
\begin{center}
\includegraphics[width=3.2in]{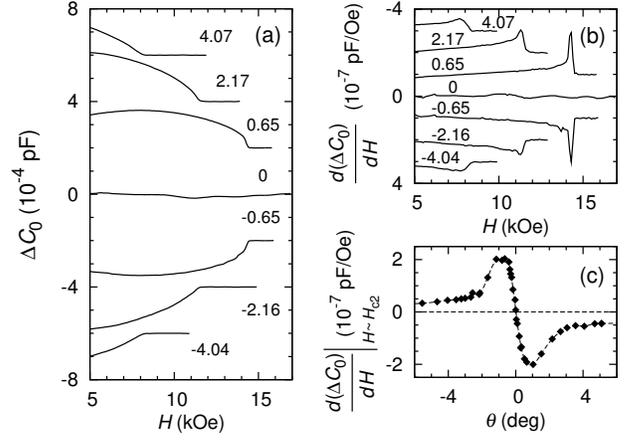}
\end{center}
\vspace{-0.2in}
\caption{
Field dependence of (a) a raw-capacitance data measured in 0~Oe/cm, $\Delta C_0$, where the normal-state value has been subtracted, and 
(b) $d(\Delta C_0)/dH$ at 0.1~K. 
Numbers labeling the curves represent the field angle $\theta$ measured from the $ab$ plane in degrees. 
Each data in (a) and (b) is vertically shifted by $\pm 2 \times 10^{-4}$~pF and $\pm 1 \times 10^{-7}$~pF/Oe, respectively, for clarity. 
(c) Angle $\theta$ dependence of the intensity of a peak in $d(\Delta C_0)/dH(H)$ appearing near $\Hc2$. 
}
\label{C0}
\end{figure}

Figures \ref{C0}(a) and \ref{C0}(b) represent the field dependence of the raw-capacitance data 
$\Delta C_0$ and $d(\Delta C_0)/dH$, respectively, measured at 0.1~K in various field orientations under a gradient field of 0~Oe/cm.
Here, the normal-state value has been subtracted for each curve.
Note that the main contribution of $\Delta C_0$ comes from the magnetic torque $\Vec{\tau}\!=\!\Vec{M} \times \Vec{H}$.
In a magnetic field exactly parallel to the $ab$ plane ($\theta\!=\!0$), $\Delta C_0(H)$ is almost invariant with changing field. 
By tilting the field orientation slightly away from the $ab$ plane, 
$\Delta C_0$ and $d(\Delta C_0)/dH$ become significantly large in the superconducting state. 
A steep change in $\Delta C_0$ and a very sharp peak in $|d(\Delta C_0)/dH|$ are seen near $\Hc2$
only when $0.2 \lesssim |\theta| \lesssim 2$~deg [Fig.~\ref{C0}(c)].
This fact, combined with the $M_{\rm av}$ jump at $\theta\!=\!0$, confirms that 
the S-N transition is of first order in a very narrow $\theta$ range of $|\theta| \lesssim 2$~deg.

\begin{figure}
\begin{center}
\includegraphics[width=3.2in]{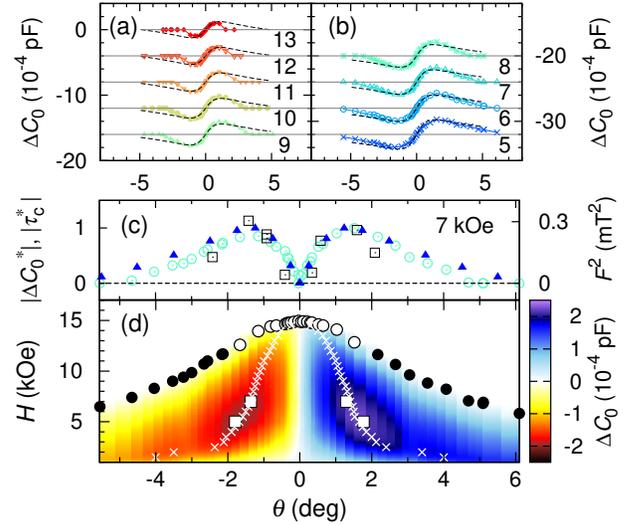}
\end{center}
\vspace{-0.2in}
\caption{
(Color online) 
(a), (b) Field-angle $\theta$ dependence of the raw-capacitance data $\Delta C_0$ at various fields for $T\!=\!0.1$~K. 
Each data is vertically shifted by $-4 \times 10^{-4}$~pF for clarity. 
Numbers labeling the curves show the applied field in kG. 
The dashed lines are the calculated results using Eq.~(1).
(c)~Angle $\theta$ dependence of $|\Delta C_0|$ 
normalized by its value at 1.5 deg, $|\Delta C_0^\ast|$, at 0.1~K (circles)
and the vortex-lattice form factor $F^2(\theta)$ at 40~mK (squares) in 7~kOe~\cite{Rastovski2013PRL}.
Triangles are the calculated data of the magnetic torque normalized by its maximum value, $|\tau^\ast_{\rm c}|$,
on the basis of the microscopic theory for a spin-singlet superconductor~\cite{Amano2014}.
The behavior for a spin-triplet superconductor with conventional orbital-limiting is expected to be essentially the same.
(d) Angle $\theta$ dependence of $\Hc2$ (circles) plotted with a contour map of $\Delta C_0(H, \theta)$. 
The open (solid) circles represent the first (second) order S-N transition.
The peak position in $|\Delta C_0(H, \theta)|$ at 0.1~K (cross), and that in $F^2(H, \theta)$ detected from SANS experiments at 40~mK (squares~\cite{Rastovski2013PRL}) are also shown.
}
\label{theta}
\end{figure}

In Figs.~\ref{theta}(a) and \ref{theta}(b), the $\Delta C_0(H,\theta)$ data at 0.1~K are plotted as a function of $\theta$ for several fixed magnetic fields.
At any fields presented here, $\Delta C_0(\theta)$ develops at low $\theta$ close to 0 deg.
In the high-field regime, e.g., 11~kOe~$\le H \le 13$~kOe, 
$\Delta C_0$ suddenly becomes zero around $|\theta| \sim 2$~deg due to the first-order S-N transition. 
By contrast, in the intermediate-field region, e.g., 5~kOe~$\le H \le 7$~kOe, 
$\Delta C_0$ remains finite and decreases gradually toward zero for $|\theta| \gtrsim 2$~deg.

The behavior in $\Delta C_0(H,\theta)$ can be understood 
as the occurrence of the transverse magnetic flux perpendicular to the applied field in a quasi-two-dimensional superconductor, irrespective of the superconducting symmetry; 
the transverse field is induced so that 
the magnetic-flux orientation is tilted toward the crystal $ab$-plane direction 
because the magnetic vortex disfavors to penetrate from one to another layer of the $ab$ plane for a small $\theta$.
The transverse flux can be detected by $\tau(\theta,H)$
as well as the vortex-lattice form factor ($F$), which reflects the spatial distribution of the transverse flux.
As represented in Fig.~\ref{theta}(d) by crosses, a peak in $|\Delta C_0(\theta)|$ always stays at $|\theta| \sim 1.5$~deg in the intermediate-field regime. 
This peak angle is in good agreement with 
that of $F^2(\theta)$ [squares in Fig.~\ref{theta}(d)] 
determined from the recent small-angle neutron scattering (SANS) experiment~\cite{Rastovski2013PRL}.
Indeed, $\Delta C_0(\theta)$ and $F^2(\theta)$ data at 7~kOe coincide sufficiently, 
as displayed in Fig.~\ref{theta}(c).
These facts support that both are attributed to the same origin, namely the induced transverse flux.

The $\theta$ dependence of $\tau$ 
for a quasi-two-dimensional superconductor with a conventional orbital-limited $\Hc2$ can be written as~\cite{Kogan2002PRL}
\begin{equation}
\tau(\theta) \propto \frac{\sin(2\theta)}{\sqrt{\cos^2\theta+\varGamma^2\sin^2\theta}}\ln\frac{\eta\varGamma H_{{\rm c2},c}}{B\sqrt{\cos^2\theta+\varGamma^2\sin^2\theta}},
\end{equation}
where $\varGamma\!=\!\xi_a/\xi_c$ is the anisotropy ratio of the coherence length, and $\eta$ is a coefficient ($\eta\!\sim\! 1$).
The peak of $\tau(\theta)$ occurring at $\theta\! \sim\! 1.3$~deg can be explained with $\eta\!=\!1.5$ and $\varGamma\!=\!60$ 
[dashed lines in Figs.~\ref{theta}(a) and \ref{theta}(b)], 
the $\varGamma_H$ value of $\SRO$ near $\Tc$~\cite{Kittaka2009JPCS}.
If we adopt $\varGamma\!=\!20$, the $\varGamma_H$ value at low temperatures~\cite{Kittaka2009PRB,Kittaka2009JPCS}, the $\tau(\theta)$ peak moves to $\theta\! \sim\! 3$~deg, in disagreement with the experiment. 
The angular variation of $\tau(\theta)$ calculated on the basis of the microscopic theory using $\varGamma=60$~\cite{Amano2014} is also in good agreement with the experiment,
as indicated by triangles in Fig.~\ref{theta}(c). 
We should note here that, although the calculation of $\tau(\theta)$ in Fig.~\ref{theta}(c) was made based on a model of spin-singlet superconductivity, it is expected that models of spin-triplet superconductivity provide nearly the same results.
These analyses suggest that the intrinsic anisotropy  $\varGamma$ of $\SRO$ is large ($\varGamma\!\sim\!60$) and independent of $T$.
This fact implies that the conventional in-plane orbital limiting field $H_{{\rm c2},ab}^{\rm orb}$ reaches $\sim\!45$~kOe at $T\!=\!0$.

To briefly summarize the experimental results, 
FOT in $\SRO$ is characterized by an entropy jump $\delta S$ of $\sim\!10$\% of the normal-state value $\gamma_{\rm n}T$~\cite{Yonezawa2013PRL}, 
a magnetization jump $\delta M$ of $\sim\!25$\% of $\chi_{\rm n}H_{{\rm c2},{ab}}$ ($\approx\!3$~G), and 
a strongly suppressed $H_{{\rm c2},ab}(0)$ ($\approx\!1/3$ of $H_{{\rm c2},ab}^{\rm orb}$). 

\begin{figure}
\begin{center}
\includegraphics[width=3.2in]{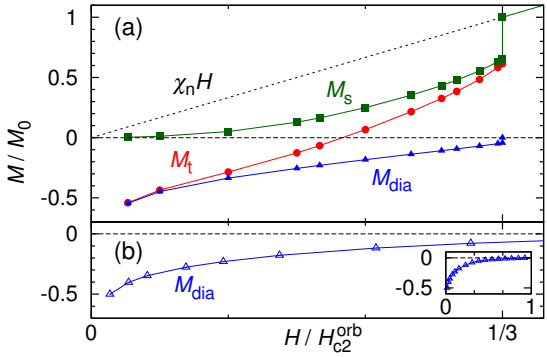}
\end{center}
\vspace{-0.2in}
\caption{
(Color online) 
(a)~Field dependence of the total magnetization $M_{\rm t}$, the spin magnetization $M_{\rm s}$, and the orbital diamagnetism $M_{\rm dia}$ 
at $T\!=\!0.1\Tc$ and $\theta\!=\!0$, obtained from the microscopic calculation 
for a Pauli-limited spin-singlet superconductor~\cite{Amano2014} with the same parameters for the calculation of $|\tau^\ast_{\rm c}|$.
Here, the calculated magnetizations are normalized by $M_0$, defined as $\chi_{\rm n}\Hc2$ in~(a).
(b)~$M_{\rm dia}$ calculated for a chiral-$p$-wave superconductor~\cite{Ishihara2013PRB} 
 with $\varGamma\!=\!60$, $\kappa\!=\!162$, $T\!=\!0.1\Tc$, and $\theta\!=\!0$, normalized by $M_0$, the same parameter in (a).
For (b), $M_{\rm s}$ shall follow $\chi_{\rm n} H$ in (a).
}
\label{calc}
\end{figure}

Note that these are similar to the characteristic features of FOT in spin-singlet SCs driven by a strong PPE. 
We calculate the field dependence of the magnetization 
of a strongly Pauli-limited spin-singlet ($s$-wave) SC at $T\!=\!0.1\Tc$ by numerically solving the microscopic Eilenberger equation 
using a three-dimensional cylindrical Fermi surface and $\varGamma\!=\!60$.
The details of the calculation method have been reported in Refs.~\onlinecite{Ichioka2007PRB} and \onlinecite{Machida2008PRB}. 
The Maki parameter $\mu$ is chosen to be 2.4 for $H\!\parallel\!ab$ and 0.04 for $H\!\parallel\!c$,
so that $H_{{\rm c2},ab}(0)/H_{{\rm c2},ab}^{\rm orb} \approx 1/3$ and $H_{{\rm c2},c}(0)/H_{{\rm c2},c}^{\rm orb} \approx 1$.
The Ginzburg-Landau parameter $\kappa\!=\!2.7$ for $H\!\parallel\!c$~\cite{Maeno2012JPSJ} is adopted and 
$\kappa$ for $H\!\parallel\!ab$ is set to be 162.
From this calculation, a clear FOT is reproducible as shown in Fig.~\ref{calc}(a), where $M_{\rm s}$ and $M_{\rm dia}$ indicate the spin and the orbital contributions to the total magnetization $M_{\rm t}$, respectively.
The jump in $M_{\rm t}$ is predominantly due to a change in $M_{\rm s}$. The diamagnetic contribution $M_{\rm dia}$ to the jump is small, roughly 10\% of that of $M_{\rm s}$.
Note that the calculated magnetizations in Fig.~\ref{calc}(a) are normalized by the value $M_0\!=\!\chi_{\rm n}\Hc2$.
If we adopt $4\pi M_0\!=\!3$~G~\cite{Maeno2012JPSJ}, the calculated $M_{\rm t}$ jump is equal to 1.1~G.
Instead, if $M_{\rm t}$ is normalized by the equality $-\int_{0}^{\Hc2} (M_{\rm t} - \chi_{\rm n}H) dH\!=\!H_{\rm c}^2/8\pi$ ($H_{\rm c}\!=\!194$~Oe~\cite{Akima1999JPSJ}), the magnetization jump becomes about 0.9~G.
In any case, the calculated discontinuity in $M_{\rm t}$ is in a reasonably good agreement with the observed value of $(0.74\pm0.15)$~G, in spite of the highly simplified model.
The slight difference between the experimental observation and the calculated $M_{\rm t}$ jump can be solved by considering the multiband effect.

However, the present results raise a fundamental quantitative issue against the PPE scenario as well.
Within the PPE scenario for spin-singlet superconductivity,
a jump in $M_{\rm s}$ as well as a jump in $S/T$ can be ascribed to a discontinuous increase in the zero-energy quasiparticle density of states.
Because of this fact, it is expected that the jump heights relative to the normal-state values in magnetization and entropy should be nearly equal to each other: 
i.e. $\delta M_{\rm s} / \chi_{\rm n}H_{{\rm c2}, ab} \simeq \delta S/\gamma_{\rm n}T$.
Indeed, a microscopic calculation supports this idea~\cite{Machida2008PRB}.
On the other hand, in the experiment, a substantial discrepancy 
between $\delta M / \chi_{\rm n}H_{{\rm c2}, ab}$ ($\sim 25$\%) and $\delta S/\gamma_{\rm n}T$ ($\sim 10$\%)~\cite{Yonezawa2013PRL} has been observed.
Hence, the observed ratio between $\delta M$ and $\delta S$ quantitatively contradicts the PPE scenario, although the Clausius-Clapeyron relation manifests the accuracy of the ratio $\delta M/\delta S$ as we described above.
In other words, $\delta M$ should contain a large fraction of non-spin contribution, and the observed $\Hc2$ slope is flatter than the expectation for the PPE scenario by a factor of 2.5.
In addition, as already mentioned, this scenario results in a sizable suppression of the spin susceptibility below $\Hc2$, 
which contradicts the NMR~\cite{Ishida2008JPCS} and neutron-scattering~\cite{Duffy2000PRL} results. 

Another question is whether the observed magnetization jump can be explained by the anticipated chiral-$p$-wave order parameter.
Microscopic calculations of the magnetization of the chiral-$p$-wave state~\cite{Ishihara2013PRB} have been done by using the parameters $\varGamma\!=\!60$ and $\kappa\!=\!162$, and an example of the results for $H\!\parallel\!ab$ at $T\!=\!0.1\Tc$ is given in Fig.~\ref{calc}(b).
Because $\Vec{H}\!\perp\!\Vec{d}$ in this configuration, the spin part $M_{\rm s}$ is irrelevant, and only the diamagnetic contribution $M_{\rm dia}$ is shown.
$M_{\rm dia}$ is suppressed smoothly toward $\Hc2$ ($=\!\Hc2^{\rm orb}$) with increasing field, and no FOT occurs as expected.
The $M_{\rm dia}$ value at  $H\!=\!\Hc2^{\rm orb}/3$, the actual upper critical field for $\SRO$, is only 0.2~G, much smaller than  the observed $M_{\rm av}$ jump of $(0.74\pm0.15)$~G.
This discrepancy can be resolved by considering the constraint  $-\int_0^{\Hc2} M_{\rm dia} dH\!=\!H_{\rm c}^2/8\pi\!=\!{\rm const}.$;
if $\Hc2$ is suppressed below the orbital limiting field by any mechanism, $M_{\rm dia}$ should be augmented so as to conserve the condensation energy.
However, at this stage, we are not aware of theoretical models to explain a strong $\Hc2$ suppression in the spin-triplet state with invariant spin susceptibility. 
Alternatively, a ``hidden'' depairing mechanism not considered in the framework of the two-dimensional chiral-$p$-wave scenario, such as those related to the internal angular moment of the Cooper pair, might be important. 
Unless such depairing mechanism is introduced, it seems difficult to reconcile the present results with the NMR and neutron Knight-shift results~\cite{Ishida2008JPCS,Ishida2014,Duffy2000PRL}.

In summary, a sharp magnetization jump of $(0.74\pm0.15)$~G, evidencing a first-order S-N transition, is clearly observed.
This result provides information toward an understanding of the superconducting nature of $\SRO$.

We thank Y. Kono for his experimental support. 
One of the authors (K. M.) thanks M. Ichioka, M. Ishihara, and Y. Amano for help with theoretical calculations, and M. R. Eskildsen for useful help in this paper.
This work was supported by KAKENHI (No. 24340075, No. 25800186, No. 22103002, No. 25103716, No. 26287078, and No. 26400360) from JSPS and MEXT.

\clearpage
\onecolumngrid
\appendix

\begin{center}
{\large Supplemental Material for \\
\textbf{Sharp magnetization jump at the first-order superconducting transition in Sr$_2$RuO$_4$}}\\
\vspace{0.1in}
Shunichiro Kittaka,$^{1}$ Akira Kasahara,$^{1}$ Toshiro Sakakibara,$^{1}$ Daisuke Shibata,$^2$ \\Shingo Yonezawa,$^2$ Yoshiteru Maeno,$^2$ Kenichi Tenya,$^3$ and Kazushige Machida$^4$\\
{\small 
\textit{$^1$Institute for Solid State Physics, University of Tokyo, Kashiwa, Chiba 277-8581, Japan}\\
\textit{$^2$Department of Physics, Kyoto University, Kyoto 606-8502, Japan}\\
\textit{$^3$Faculty of Education, Shinshu University, Nagano 310-8512, Japan}\\
\textit{$^4$Department of Physics, Okayama University, Okayama 700-8530, Japan}\\
}
(Dated: \today)
\end{center}

\section{I. Details of the experimental setup}
Magnetization measurements at temperatures down to 0.1~K were performed by a capacitively-detected Faraday method in a dilution refrigerator.
A translational magnetic force ($M dH_z/dz$) acting on a magnetic moment $M$ situated in a spatially-varying field $H_z(z)$ was detected by a transducer made of a parallel-plate capacitor. 
One of the capacitor plate, on which a sample was mounted, was suspended by thin phosphor-bronze wires and  could move in proportion to an applied force. 
Hence, the magnetic force was detected by a change in the capacitance value $\Delta C$.
For this purpose, we used a vertical superconducting solenoid equipped with a pair of gradient coils, 
driven by  independent power supplies; the field gradient ($dH_z/dz$) at the sample position could be varied independent of the central field [$H_z(0)$].
For a magnetically anisotropic sample, a torque component (${\pmb M}\times{\pmb H}$)  is usually superposed on the capacitor output. 
In order to eliminate the torque effect and to obtain the magnetization, we took a difference between the capacitance data with $dH_z/dz\neq 0$ and those ($\Delta C_0$) with $dH_z/dz= 0$, the latter providing the torque component only. The zero-gradient capacitance data $\Delta C_0$ were also used to analyze the field variation of the magnetic torque.

To measure the magnetization and the magnetic torque under a precise control of the magnetic-field orientation, 
we developed a device illustrated in Fig. S1.
The capacitor transducer is mounted on a tilting stage, which can be rotated around the $x$ axis. 
The tilting angle is adjusted by rotating a screw rod from the top of the dilution insert using an upper shaft made of glass epoxy, 
which goes through a line-of-sight port of the refrigerator insert. One revolution of the screw rod corresponds to a rotation of the tilting stage of 1.5 deg [Fig. S1 (b)].
The revolution of the screw rod is read by a potentiometer dial.
The sample is mounted on a sample stage of the capacitor transducer so that the [110] axis coincides with the rotational axis of the tilting stage.

In order to cut a heat flow through the upper shaft into the capacitor transducer, we use a thermal isolator as illustrated in Fig. S1 (c).
When the upper shaft is rotated, two arms touch the columns and transmit revolution to the lower shaft. 
During the measurement, 
the arms and the columns are detached so that the upper and lower shafts are thermally isolated to each other.
Then, the sample temperature can reach below 0.1~K.
In order to avoid a backlash of the isolator, we always read the potentiometer dial with a clockwise rotation.

We also improved the sensitivity of the magnetization measurement by a factor of 100 over the previous apparatus used in Ref.~24,
by reducing the mass of the movable capacitor plate, on which a sample was mounted, and making the background magnetization significantly smaller.

\section{II. Temperature variation of the magnetization curve}
Figure~S2(a) shows the field dependence of the magnetization $M_{\rm SC}\!=\!M\!-\!\chi_{\rm n}H$ of $\SRO$
taken in the increasing and decreasing field sweeps ($M_{\rm SC}^{\rm u}$ and $M_{\rm SC}^{\rm d}$, respectively), at various temperatures for $H \parallel ab$.
We also plot a field derivative $dM_{\rm SC}/dH$ of the increasing-field data, $dM_{\rm SC}^{\rm u}/dH$, in Fig.~S2(b).
The ramp rate of the magnetic field is 750 Oe/min in all the measurements, and each data point is taken while the field is held constant.
With increasing temperature, the magnetization jump and the peak height of $dM_{\rm SC}^{\rm u}/dH$ become smaller and broader.
Above 0.6 K, the amplitude of the $dM_{\rm SC}^{\rm u}/dH$ peak is strongly suppressed [see Fig. S2(c)].
The present results demonstrate that the first-order S-N transition becomes remarkable for $T \lesssim 0.6$~K.

\setcounter{figure}{0}
\renewcommand{\thefigure}{S\arabic{figure}}
\begin{figure}
\includegraphics[width=6.in]{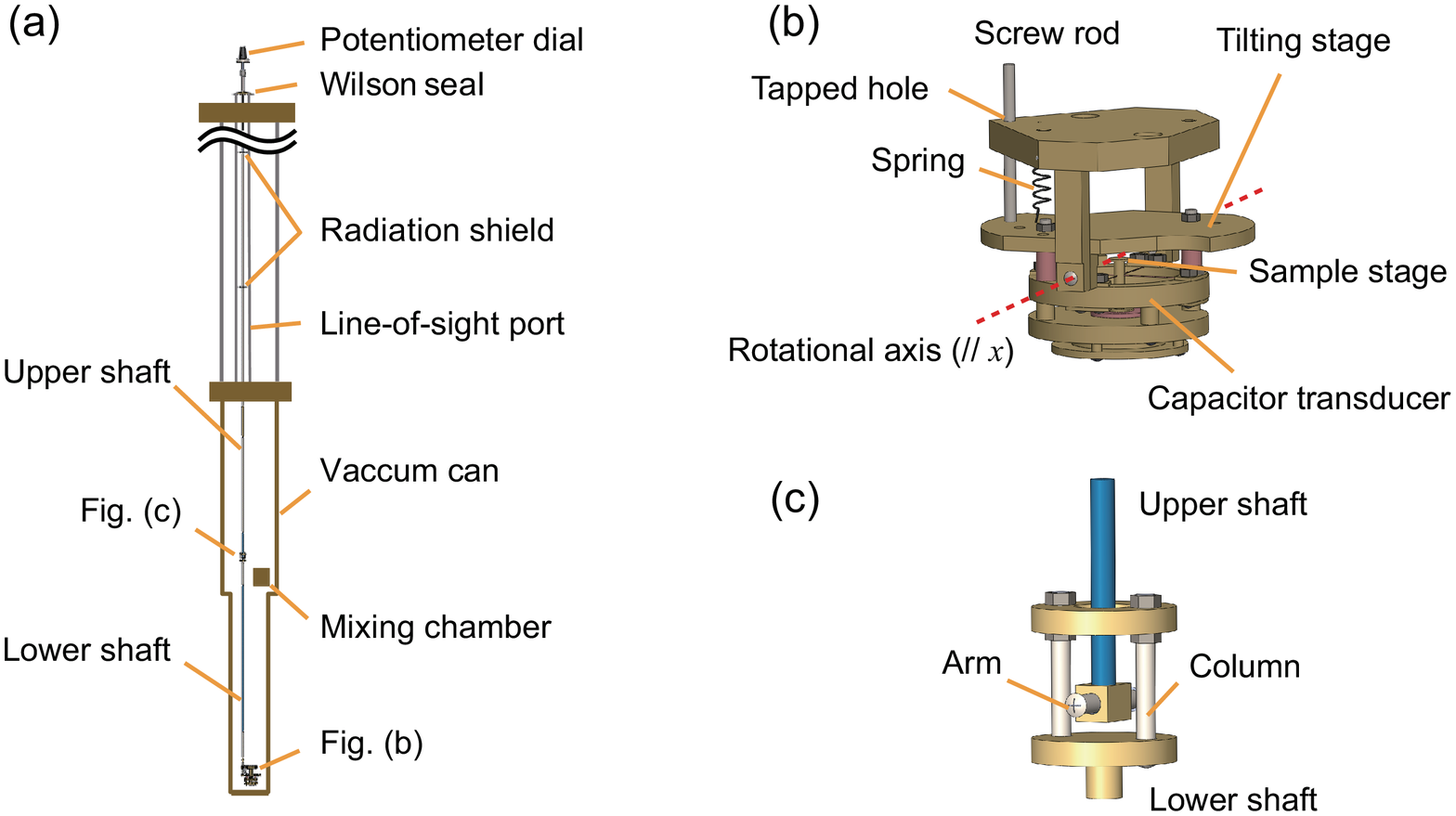} 
\caption{
(Color online) 
(a) Schematic view of the device for fine tuning of the field angle $\theta$, on which a capacitor transducer is mounted.
(b) Enlarged view of the tilting stage and a capacitor transducer. The sample was fixed on the sample stage at $z=0$ and 
was rotated around the $x$ axis by rotating the tilting stage. 
One side of the tilting stage is pulled up (pushed down) by the spring (the screw rod connected to the lower shaft).
(c) Enlarged view of the thermal isolator to cut the heat flow. 
The arms and columns were touched (detached) during changing $\theta$ (the measurement).
}
\label{Supp}
\end{figure}

\begin{figure}[b]
\begin{center}
\includegraphics[width=6.3in]{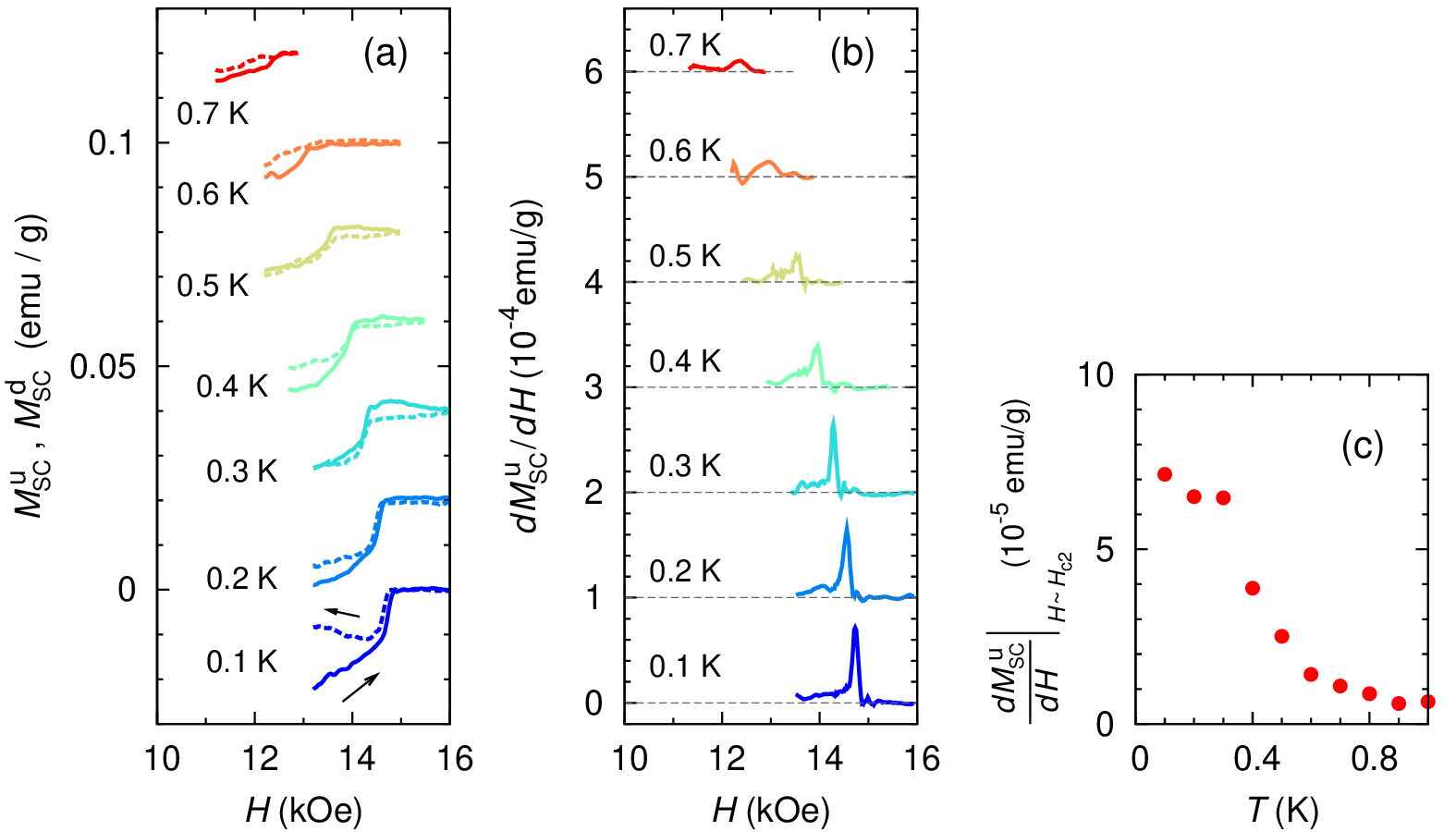}
\end{center}
\caption{
(Color online) 
Field dependence of (a) $M_{\rm SC}^{\rm u}$ (solid line), $M_{\rm SC}^{\rm d}$ (dashed line), and (b) $dM_{\rm SC}^{\rm u}/dH$ at various temperatures.
Each data in (a) and (b) is vertically shifted by 0.02 and $1 \times 10^{-4}$~emu/g, respectively, for clarity.
(c) Temperature dependence of the peak height in $dM_{\rm SC}^{\rm u}/dH(H)$ appearing near $\Hc2$. 
}
\label{fig2}
\end{figure}

\end{document}